\documentclass[11pt, epsf]{article}
\usepackage{epsfig}
\def\ga{\mathrel{\raise.3ex\hbox{$>$\kern-.75em\lower1ex\hbox{$\sim$}}}}
\def\la{\mathrel{\raise.3ex\hbox{$<$\kern-.75em\lower1ex\hbox{$\sim$}}}}

\def\I_M{{I_{\scriptscriptstyle M\times M}}}

\begin{document}

\thispagestyle{empty}

\vskip 2cm \centerline{ \Large \bf  Tunneling of Massive Vector Particles from Dilaton Black Holes}
\vskip .2cm
\centerline{\Large\bf  in $(2+1)$ Dimensions}

\vskip .2cm

\vskip 1.2cm

\centerline{ \bf Anindya Biswas\footnote{Electronic address: 
ani\_imsc@yahoo.co.in}}
\vskip 10mm \centerline{ \it Department of Physics,}
 \vskip 5mm
\centerline {Ranaghat College, Ranaghat, India.} 
\vskip 1.2cm
\vskip 1.2cm
\centerline{\bf Abstract}
\noindent
In this paper, we have studied the Hawking radiation of massive spin-$1$ particles from the black holes in $(2+1)$ dimensions with non- trivial dilaton fields. We consider two special varities of these black holes one is static charged and other is spinning electrically neutral. By applying the standard method of $WKB$ approximation and Hamilton- Jacobi ansatz we have shown the tunneling probability and Hawking temperature of massive bosons accordingly. In the certain limit of the dilaton coupling for spinning neutral case we have recovered the Hawking temperature of the $BTZ$ black holes as well.     

\newpage
\setcounter{footnote}{0}
\noindent
\section{Introduction}
Black holes have become the most fascinating objects to study in the context of Einstein's general theory of relativity over the last century. According to Hawking \cite{Hawk11}, \cite{Hawk22}, \cite{Hawk33} these black holes emit thermal radiation at a constant temperature across the horizon. This temperature is known as Hawking temperature. Several method have been developed to understand Hawking radiation and calculate Hawking temperature after the original method proposed by Hawking \cite{Hawk11}. In recent years semi-classical approach \cite{Srini}-\cite{Wang} of deriving Hawking radiation as tunneling process captured lot of attention. 
This semi- classical tunnelling method is considered to calculate the imaginary
part of the action for the (classically forbidden) process of s-wave emission
across the horizon (Kraus and Wilczek \cite{Krau11}- \cite{Krau33}), which
is related to the Boltzmann factor for emission at the Hawking temperature.
Applying the WKB approximation the emission and absoprption probabilities for the classically forbidden trajectory of the s-wave coming from inside to outside of the horizon is
given by, $\Gamma \simeq exp(-2ImI)$, where $I$ is the classical action of the trajectory of the tunneling particles.   
In calculating the Hawking temperature two different methods were developed to evaluate the imaginary part of the action- one was proposed by Parikh and Wilczek \cite{Parikh} and the other method used by Angheben et.al. \cite{Angheben}, this is an extension of the analysis done by Srinivasan and Padmanabhan \cite{Srini}. \\
First method deals with the part of the action that contributes an imaginary term is 
$ImI=\int_{r_{in}}^{r_{out}}{p_rdr}$, where $p_r$ is the emitted null $s$-wave. Hence one can calculate imaginary part of the action by using the Hamilton's equation and the knowledge of null geodesics. This method is known as {\it null geodesic method}. The second method is known as the {\it Hamilton-Jacobi method} where the imaginary part of the action is derived from the Hamilton-Jacobi equation by choosing suitably the form of the action. This tunneling method as well applied for the black hole radiation of spin-$\frac{1}{2}$ and spin-$\frac{3}{2}$ particles and photons at Hawking temperature \cite{Kerner11}, \cite{Kerner22}, \cite{Yale}, \cite{Jing}, \cite{Majhi}, \cite{Yale11} close to the black hole horizon.        \\
In continuation of using further the tunneling method to model spin-$1$ 
(e.g. bosons like $W^\pm$ and $Z$) particle emission from black holes S. I. Kruglov \cite{Krug} investigated Hawking radiation in $(2+1)$ dimensions. Recently various authors have analysed Hawking radiation of vector bosons in the context of static and rotating black holes \cite{Krug}- \cite{Jusufi}. Apart form the various black hole solutions with or without cosmological constants here we are interested black holes with nontrivial dilaton fields which are neither asymptotically flat nor anti-de Sitter. We are interested to investigate Hawking radiation and to calculate Hawking temperature of a class of Einstein-Maxwell dilaton black holes with arbitrary dilaton coupling as discovered by Chan and Mann \cite{ChanMann11}, \cite{ChanMann22}.
Therefore our purpose is to study the tunneling of massive vector boson $W^\pm$ through the space- time of a static charged \cite{ChanMann11} and spinning charge neutral \cite{ChanMann22} dilaton black holes in $(2+1)$ dimensions. In case of static charged black holes we derive the field equation of the charged boson from the Lagrangian given by Glashow-Weinberg-Salam model. After that we use the WKB approximation and then apply the Hamilton- Jacobi ansatz, for solving the radial part of the action by using the determinant of the matrix equals zero, we found the tunneling rate and the corresponding Hawking temperature. For the later case of spinning black hole we use Proca equation as the field equation of the massive vector field which is neutral. We again calculate the tunneling rate in case of spinning dilaton black holes and also recover associated Hawking temperature of this class of black holes. We are mainly interested to study Hawking radiation in the background of these special class of black holes for generic values of dilaton coupling. We have also obtained two known class of black holes like 
$BTZ$ \cite{Brown} and $MSW${\footnote{elaborate discussion on the $MSW$ black holes are given in \cite{ChanMann11} and in the original paper \cite{MSW}}} in the two special limits of the dilaton coupling and also can reproduce the Hawking temperature for those particular class of black holes.\\
This paper is organised as follows. In section $2$ we consider static charged black holes with nonzero dilaton filed and investigate the Hawking radiation of charged vector bosons and estimate the Hawking temperature. In section $3$ we also study tunneling of massive vector bosons which are uncharged in the spinning neutral dilaton black hole space- time. We conclude our study in section $4$.               
\section{Tunneling from Static Charged Dilaton Black Holes}
We consider the action given in {\cite{ChanMann11}} as follows
\begin{equation}
I = \int d^3x \sqrt{-g}(\mathcal{R}- 4(\nabla\phi)^2+ 2 e^{b\phi}{\Lambda}- 
e^{-4a\phi}F^2)
\label{action11}
\end{equation}
with arbitrary couplings $a$ and $b$. Here $\mathcal{R}$ is the Ricci scalar, 
$\phi$ is the dilaton field, $\Lambda$ is the cosmological constant and $F$ is the Maxwell's field strength. When $a=1$ and $b=4$, the 
eqn.(\ref{action11}) is a action for low energy string theory expressed in a usual einstein metric. Now the family of static circularly symmetric black hole solutions to the above action as obtained in {\cite{ChanMann11}} is
\begin{equation}
ds^2=-f(r)dt^2 + h(r)^{-1}{dr^2} + r^2 d\theta^2,
\label{solt11}
\end{equation}
where
\begin{equation}
f(r)= -\frac{2M}{N}r^{\frac{2}{N}-1} + \frac{8\Lambda}{(3N-2)N}r^2+\frac{8Q^2}{(2-N)N}, 
\label{metric11}
\end{equation}
and
\begin{equation}
h(r)^{-1} = \frac{4r^{\frac{4}{N}-2}}{N^2\gamma^{\frac{4}{N}}}{f(r)^{-1}}.
\label{metric22}
\end{equation}
According to \cite{ChanMann11} $M$ is the quasilocal mass. 
The corresponding dilaton field $\phi$ and the non-zero electric field $F_{tr}$ are 
\begin{eqnarray}
\phi &=& \frac{2k}{N}ln\Big(\frac{r}{\beta(\gamma)}\Big), \nonumber \\
F_{tr} &=& \frac{Qe^{4a\phi}}{r}, \nonumber \\
k &=& \pm \frac{1}{4}\sqrt{N(2-N)}, \nonumber \\
4ak &=& bk = N-2,
\label{Multy}
\end{eqnarray}

$Q$ is related to the electric charge whereas $\beta$, $\gamma$ are integration constants. $k$ is a real number. Last relations in eqn.(\ref{Multy}) are among the various dilaton couplings, finally related with a single parameter $N$. 
As $(M,\Lambda)> 0$ the black hole solutions exist only in the domain $2\geq N> \frac{2}{3}$. When $N=1$, (\ref{solt11}) reduces to the $2+1$ dimensional $MSW$. For $N=2$ and $Q=0$, metric (\ref{solt11}) represents uncharged $BTZ$ black hole which might be the special case of $BTZ$ black hole given in {\cite{BTZ}}. In general due to the presence of nontrivial dilaton, black holes are neither asymptotically flat nor anti-de Sitter (de Sitter).\\\\
Let us now consider the Lagrangian of $W$-bosons which is treated as massive vector boson here, interact with the background electromagnetic field is given by {\cite{LiChen}},
\begin{equation}
\mathcal{L} = - {\frac{1}{2}}(D_\mu^+ W_\nu^+ - D_\nu^+ W_\mu^+)(D^{-\mu}W^{-\nu} - D^{-\nu}W^{-\mu})+\frac{m^2}{\hbar^2}W_\mu^+W^{-\mu}-\frac{i}{\hbar}eF^{\mu\nu}W_\mu^+ W_\nu^-
\label{LagGWS}
\end{equation}
where $D_\mu^{\pm} = \nabla_\mu\pm\frac{i}{\hbar}eA_\mu$ and $\nabla_\mu$ is the covariant geometric derivative. $A_\mu$ is the vector potential associated with the electromagnetic field $F^{\mu\nu}$ (where $F^{\mu\nu} = \nabla^{\mu}A^\nu-\nabla^{\nu}A^{\mu}$) of the black hole. The charge carries by the $W^+$ boson is denoted by $e$ and the mass is $m$. Equation of motion of the $W$-boson should be derived from the above Lagrangian and write down in the following form
\begin{equation}
\frac{1}{\sqrt{-g}} {\partial_\mu}[\sqrt{-g}(D^{\pm\nu}W^{\pm\mu}-D^{\pm\mu}W^{\pm\nu})]\pm
\frac{ieA_\mu}{\hbar}(D^{\pm\nu}W^{\pm\mu}-D^{\pm\mu}W^{\pm\nu})+
\frac{m^2}{\hbar^2}W^{\pm\nu}\pm\frac{i}{\hbar}eF^{\mu\nu}W_\mu^{\pm}=0
\end{equation}
In order to study the tunneling of $W^+$ bosons as the spin-$1$ particle, one should solve the following equation
\begin{eqnarray}
\frac{1}{\sqrt{-g}}\partial_\mu[\sqrt{-g}g^{\mu\alpha}g^{\nu\beta}(\partial_\beta W^{+}_\alpha-\partial_\alpha W_\beta^+ + \frac{i}{\hbar}eA_\beta W^+_\alpha-\frac{i}{\hbar}eA_\alpha W^+_\beta)] \nonumber \\
+\frac{ieA_\mu g^{\mu\alpha}g^{\nu\beta}}{\hbar}(\partial_\beta W^{+}_\alpha-\partial_\alpha W_\beta^+ + \frac{i}{\hbar}eA_\beta W^+_\alpha-\frac{i}{\hbar}eA_\alpha W^+_\beta) + \frac{m^2}{\hbar^2}g^{\nu\beta}W_\beta^+ + \frac{i}{\hbar}eF^{\nu\lambda}W_\alpha^+=0
\label{EQM11}
\end{eqnarray}
The solution of eqn.(\ref{EQM11}) can be represented in the form 
\begin{equation}
W_\mu^+(t,r,\theta)= C_\mu e^{\frac{i}{\hbar}S(t,r,\theta)},
\label{Wmu}
\end{equation}
where $\mu= 0,1,2$ and $C_\mu$ represents some arbitrary functions.
According to the WKB approximation, one can write down the action as
\begin{equation}
S(t,r,\theta)= S_0(t,r,\theta)+\hbar S_1(t,r,\theta)+\hbar^2 S_2(t,r,\theta)+......
\label{Action11}
\end{equation}
The determinant of the metric in eqn.(\ref{solt11})
\begin{equation}
g=-{r^2} \frac{f}{h},
\label{Deter}
\end{equation}
and the nonzero components of the metric are
\begin{equation}
g^{00}=-\frac{1}{f},~~~~   g^{11}=h,~~~~      g^{22}=\frac{1}{r^2}.       
\label{Inverse}
\end{equation}
Using eqns(\ref{EQM11}), (\ref{Wmu}), (\ref{Action11}), (\ref{Deter}) and (\ref{Inverse}) we can get three sets of equations in the leading order of   
$\hbar${\footnote {$h$ and $\hbar$ denote two different quantity}}
\begin{equation}
C_0[(\partial_1 S_0)^2+\frac{1}{r^2 h}(\partial_2 S_0)^2 + \frac{m^2}{h}]-C_1[(\partial_1S_0)(\partial_1S_0 + eA_0)]- \frac{C_2}{r^2 h}[(\partial_2S_0)(\partial_0S_0+eA_0)]=0,
\label{matrix1}
\end{equation}
\begin{equation}
C_0[-(\partial_1S_0)(\partial_0S_0+eA_0)]+ C_1[-\frac{f}{r^2}(\partial_2S_0)^2+(\partial_0S_0+eA_0)^2-m^2f]+C_2\frac{f}{r^2}(\partial_2S_0)(\partial_1S_0)=0,
\label{matrix2}
\end{equation}
\begin{equation}
C_0[(\partial_2S_0)(\partial_0S_0+eA_0)]-hfC_1(\partial_1S_0)(\partial_2S_0)-C_2[(\partial_0S_0+eA_0)^2-m^2f-hf(\partial_1S_0)^2]=0.
\label{matrix3}
\end{equation}
We assume that the action $S_0(t,r,\theta)$ can be written in the form
\begin{equation}
S_0(t,r,\theta)=-Et+W(r)+J\theta+K
\label{S_0}
\end{equation}
where $E$ and $J$ are the energy and the angular momentum of the $W^+$ boson respectively. Whereas $K$ is a complex constant.
Inserting eqn.({\ref{S_0}}) into eqns. (\ref{matrix1})- (\ref{matrix3}), one can obtain a matrix equation $\Xi(C_0,C_1,C_2)^T=0$ (here $T$ denotes the transition to the transposed vector)
where $\Xi$ is a $3\times3$ matrix whose elements are
\begin{eqnarray}
\Xi_{11} &=& W^{\prime2}+\frac{J^2}{r^2 h}+\frac{m^2}{h}   \nonumber\\
\Xi_{12} &=& W^\prime(E-eA_0)     \nonumber\\
\Xi_{13} &=& \frac{J}{r^2h}(E-eA_0)     \nonumber\\
\Xi_{21} &=& W^\prime(E-eA_0)      \nonumber\\
\Xi_{22} &=& -\frac{f}{r^2}J^2+(E-eA_0)^2-m^2f      \nonumber\\
\Xi_{23} &=& \frac{f}{r^2}JW^\prime       \nonumber\\
\Xi_{31} &=& J(E-eA_0)        \nonumber \\
\Xi_{32} &=& hfJW^\prime         \nonumber \\
\Xi_{33} &=& (E-eA_0)^2-m^2f-hfW^{\prime2} 
\label{Lambda}        
\end{eqnarray}
To obtain the nontrivial solution of the Eqns. (\ref{matrix1}-\ref{matrix3}), we should have
the determinant of the matrix $\Xi$ equals to zero. So by solving $det\Xi=0$ we get
\begin{equation}
[(E-eA_0)^2-r^2fhW^{\prime2}-(m^2r^2+J^2)f]^2=0.
\end{equation}
Solving this equation for the radial function $W(r)$ we have the following integration
\begin{equation}
W_\pm=\pm\sqrt{\frac{(E-eA_0)^2-f(r)(\frac{J^2}{r^2}+m^2)}{h(r)f(r)}}.
\end{equation}
Integrating around the pole at the event horizon $r_+$ we obtain
\begin{equation}
W_\pm= \pm {i\pi}\frac{(E-eA_0)}{\sqrt{h^\prime{(r_+)}f^\prime{(r_+)}}}.
\end{equation}
$W_+$ denotes the radial function of outgoing particles and $W_-$  of the ingoing particles. Then following \cite{Kerner33} we can obtain the tunnelling probability of $W^+$ bosons   
\begin{equation}
\Gamma \simeq e^{(-4Im W_+)}.
\label{Gamma}
\end{equation}
We can find the Hawking temperature by simply comparing eqn.(\ref{Gamma}) with the Boltzmann factor $\Gamma=e^{\beta E_{net}}$, where $E_{net} = (E-eA_0)$ and $\beta = 1/T_H$, producing 
\begin{equation}
T_H = \frac{1}{4\pi}\sqrt{f^\prime{(r_+)}h^\prime{(r_+)}}.
\label{Hawktemp11}
\end{equation}
Using Eqns.(\ref{metric11}), (\ref{Hawktemp11}), one can recover the Hawking temperature for a static charged dilaton black hole as derived in \cite{ChanMann11}
\begin{equation}
T_H= \frac{\lambda \gamma^{{2}\over{N}}M}{2\pi r_+}\Big[\Big(\frac{r_+^2}{r_m^2}\Big)^{1-\lambda}-1\Big],
\label{Temp}
\end{equation}
where $\lambda = {{2-N}\over{2N}}$ and $r_m=\Big(\frac{(3N-2)(2-N)M}{8N\Lambda}\Big)^{\frac{N}{3N-2}}$. Here the location of the outer(event) horizon $r_+$ is a function of the black hole parameters $M$, $Q$, $\Lambda$ and dilaton coupling $N$.
We can show for uncharged case (i.e. $Q=0$) Hawking temperature becomes 
\begin{equation}
T_H= \frac{\gamma^\frac{2}{N}\Lambda}{\pi N}\Big(\frac{(3N-2)M}{4\Lambda}\Big)^{\frac{2(N-1)}{3N-2}}.
\label{T_HQ=0}
\end{equation} 
Following our earlier discussion the Hawking temperature derived in eqn.(\ref{Temp}) reduces to the Hawking temperature $T_H= \frac{\gamma^2 M}{4\pi r_+}\sqrt{1-\frac{64Q^2\Lambda}{M^2}}$ for charged $MSW$ black hole for $N=1$ which is exactly matched with the result presented in \cite{Sakalli44}. Furthermore for vanishing $Q$ we can get from eqn.(\ref{T_HQ=0}) for uncharged static $BTZ$ black hole $T_H=\frac{\gamma}{2\pi}\sqrt{M\Lambda}$, for uncharged $MSW$ black hole the temperature would be $T_H=\frac{\gamma^2\Lambda}{\pi}$, which only depends on the cosmological constant $\Lambda$.
\section{Tunnelling from Spinning Dilaton Black Holes.}
We mention below a class of spinning black hole solutions parametraised by the mass $M$, angular momentum $J$ and dilaton coupling $N$ which metric is given by \cite{ChanMann22} as
\begin{equation}
ds^2=-f(r)dt^2+\frac{dr^2}{h(r)}+2l(r)dtd\theta+p(r)^2d\theta^2
\label{metric22}
\end{equation}
with
\begin{equation}
f(r)=[\frac{8\Lambda r^N}{(3N-2)N}+\delta r^{1-{{N}\over{2}}}],
\end{equation}
\begin{equation}
h(r)=\Big[\frac{8\Lambda r^N}{(3N-2)N}+\Big(\delta-\frac{2\Lambda\omega^2}{(3N-2)N\delta}\Big)r^{1-{N}\over{2}}\Big],
\end{equation}
\begin{equation}
l(r)=-\frac{\omega r^{1-{{N}\over{2}}}}{2},
\end{equation}
\begin{equation}
p(r)^2=\Big[r^N-\frac{\omega^2}{4\delta}r^{1-{N}\over{2}}\Big].
\label{pr}
\end{equation}
The mass $M$ and angular momentum $J$ are related with the parameters $\delta$ and $\omega$ as follows
\begin{equation}
M= {{N}\over{2}}\Big[\frac{2\Lambda\omega^2}{(3N-2)N\delta}\Big({{4}\over{N}}-3\Big)-\delta\Big],
\end{equation}
\begin{equation}
J=\frac{3N-2}{4}\omega,
\end{equation}
\begin{equation}
\delta=-{{M}\over{N}}-\sqrt{\frac{M^2}{N^2}+\Big(\frac{4}{N}-3\Big)\frac{2\Lambda\omega^2}{(3N-2)N}}.
\end{equation}
As described by Kruglov \cite{Krug} and subsequently by others in \cite{CZH}, the Proca equation for the massive vector particles having the wave function $\psi$ is given by
\begin{equation}
\frac{1}{\sqrt{-g}}\partial_\mu(\sqrt{-g}\psi^{\nu\mu})+\frac{m^2}{\hbar^2}\psi^\nu=0,
\label{Proca}
\end{equation}
where
\begin{equation}
\psi_{\mu\nu}=\partial_\mu\psi_\nu-\partial_\mu\psi_\nu.
\label{Psimu}
\end{equation}
The solution of the eqn.(\ref{Proca}) after using the form of the metric given in eqn.(\ref{metric22}) is considered in the form
\begin{equation}
\psi_\mu(t,r,\theta)= C_\mu e^{\frac{i}{\hbar}S(t,r,\theta)},
\label{Psi}
\end{equation}
again $\mu= 0, 1, 2$.
With the help of the WKB approximation, one can represent the action similar to 
eqn.(\ref{Action11}) as
\begin{equation}
S(t, r, \theta) = S_0(t, r, \theta) + \hbar S_1(t, r, \theta) + \hbar^2 S_2(t, r, \theta) + ....
\label{Act}
\end{equation}
However we further choose the action in the form  
\begin{equation}
S_0= −Et + W(r) + j\theta + K,
\label{Action22}
\end{equation}
where $E$ and $j$ are the energy and angular momentum of the spin-1 particles, respectively,
and $K$ is a (complex) constant. After substituting Eqns. (\ref{Psimu}), (\ref{Psi}), (\ref{Act}) and (\ref{Action22}) into eqn.(\ref{Proca}) and
keeping only the leading order term in $\hbar$, we obtain a $3\times3$ matrix (matrix is $\Xi$) equation: $ \Xi(C_0, C_1, C_2)^T = 0$. Thus, one can read the non-zero components of $\Xi$ as follows
\begin{eqnarray}
\Xi_{11} &=& -p^2hW^{\prime2}-J^2-p^2m^2, ~~\Xi_{12}= -p^2hEW^\prime-hgJW^\prime \nonumber \\\
\Xi_{13} &=& hgW^{\prime2}-EJ+gm^2, ~~\Xi_{21}= -p^2hEW^\prime-hgJW^\prime \nonumber \\\
\Xi_{22} &=& -p^2hE^2-2hgEJ+fhJ^2+h(fp^2+g^2)m^2,~\Xi_{23}= hgEW^\prime-fhJW^\prime \nonumber \\\
\Xi_{31} &=& -EJ+hgW^{\prime2}+m^2g,~~ \Xi_{32}=hgEW^\prime-hfJW^\prime,\nonumber \\\
\Xi_{33} &=&-E^2+hfW^{\prime2}+m^2f.
\label{Matrix22}
\end{eqnarray}
The system of linear equations $\Xi(C_0, C_1, C_2)^T = 0$ possesses nontrivial solutions if $det\Xi = 0$. Then from eqn.(\ref{Matrix22}) we obtain
\begin{equation}
-hm^2[-2gEJ+fJ^2+g^2\{m^2+hW^{\prime2}\}+p^2\{-E^2+f(m^2+hW^{\prime2})\}]^2=0.
\label{Eqn12}
\end{equation}
From eqn.(\ref{Eqn12}) we get
\begin{equation}
W^{\prime2}={{p^2E^2+2gEJ-fJ^2-m^2(g^2+fp^2)}\over{h(g^2+p^2f)}}.
\end{equation}
The above equation immediately gives the function $W(r)$ as
\begin{equation}
W_\pm=\pm\int{\frac{\sqrt{-\frac{(3N-2)N\delta}{8\Lambda}r^{1-\frac{N}{2}}(E-\Omega_+ j)^2+h(r)\Big(\frac{(3N-2)N}{8\Lambda r^N}-m^2-\frac{j^2}{r^N}\Big)}}{r^\frac{N}{2}h(r)}},
\label{Wfun}
\end{equation}
where $\Omega_+=-\frac{\omega {r_+}^{1-\frac{N}{2}}}{2{p^2(r_+)}}$ is angular velocity at the outer event horizon located at $r_+$. We mention that $W_+$ denotes the radial function of the outgoing particles and 
$W_-$ of the ingoing particles. Integrating around the pole at the outer horizon, we obtain
\begin{equation}
W_\pm=\pm i\pi \frac{\sqrt{-\frac{(3N-2)N\delta}{8\Lambda}r_+^{1-\frac{3N}{2}}}}{h^\prime(r_+)}(E-\Omega_+j),
\label{Wfun12}
\end{equation}
As previously we calculate the tunneling probability of the massive vector particle is
\begin{equation}
\Gamma \simeq e^{(-4Im W_+)}.
\end{equation}
We have found in eqn.(\ref{Wfun12}), the imaginary part of the function $W(r)$ is
\begin{equation}
ImW_+=\frac{2\pi}{\kappa}(E-\Omega_+j).
\end{equation}
Here $\kappa$ is defined as the surface gravity of the black hole horizon and the Hawking temperature would be
\begin{equation} 
T_H= \frac{\kappa}{2\pi}=\frac{\Lambda}{\pi N}{{r_+^{\frac{3N}{2}-1}\over{p(r_+)}}}= \frac{N}{4\pi p(r_+)}(\frac{3N-2}{4-3N})\Big[\frac{M}{N}(\frac{2}{N}-1)+\sqrt{\frac{4M^2}{N^2}+(\frac{4}{N}-3)\frac{8\Lambda\omega^2}{(3N-2)N}}(\frac{1}{N}-1)\Big].
\label{Hawtemp22}
\end{equation}
We have put the expression for the location of the outer event horizon $r_+$ by setting the function $h(r_+)=0$ from the metric (\ref{metric22}) as
\begin{equation}
r_+^{\frac{3N}{2}-1}=\frac{(3N-2)N^2}{4\Lambda(4-3N)}\Big[\frac{M}{N}(\frac{2}{N}-1)+\sqrt{\frac{4M^2}{N^2}+(\frac{4}{N}-3)\frac{8\Lambda\omega^2}{(3N-2)N}}(\frac{1}{N}-1)\Big].
\end{equation}
We get the Hawking temperature in eqn.(\ref{Hawtemp22}) exactly same as the standard Hawking temperature derived in \cite{ChanMann22}. If the angular momentum J 
(or spinning parameter $\omega$) approaches zero, the metric (\ref{metric22}) reach the metric of the non-spinning dilaton black hole discussed in the previous section and the Hawking temperature (\ref{Hawtemp22}) nicely go back to the temperature of the uncharged nonspinning black hole shown in eqn.(\ref{T_HQ=0}). 
Here we also present the expression for the Hawking temperature for two limits of the dilaton coupling parameter $N$. The first one is the spinning generalization of the $MSW$ black hole in $2+1$ dimensions which temperature can be obtained from eqn.(\ref{Hawtemp22}) by simply putting the coupling parameter $N=1$ as follows:  
\begin{equation}
T_H=\frac{M}{4\pi p(r_+)}= \frac{1}{4\pi}\Big[\frac{1}{16\Lambda^2}+\frac{J^2}{M\Lambda(M+\sqrt{M^2+32\Lambda J^2})}\Big]^{-\frac{1}{2}}.
\end{equation}
Secondly for $N=2$ the metric (\ref{metric22}) after suitable coordiante transformation reduces to the metric of spinning $BTZ$ black hole and the Hawking temperature of $BTZ$ black hole is produced immediately as obtained in \cite{Brown} is
\begin{equation}  
T_H=\frac{\Lambda}{2\pi p(r_+)}\sqrt{M^2-\Lambda J^2}=\frac{\Lambda}{2\pi p_+}(p_+^2-p_-^2),
\end{equation}
here $p_+=p(r_+)$ and $p_-=p(r_-)$ are the location of outer and inner event horizon respectively for the $BTZ$ black hole.  

\section{Conclusion}
In this study we have presented the semi-classical tunneling processes to understand Hawking radiation of massive vector particles either charged or uncharged, tunneling through the event horizon of the particular class of dilaton black holes in $(2+1)$ dimensions. We have computed first the probability of tunneling of charged spin-$1$ ($W^+$ bosons) particles through the event horizon of static charged black hole which has non-trivial dilaton field. Further we have calculated Hawking temperature at horizon of this charged black hole for arbitrary coupling parameter $N$. This results exactly matched with the Hawking temperature obatined by the other method in references \cite{ChanMann11} for arbitrary coupling $N$, and in \cite{Sakalli44} for $N=1$. In the latter part of the paper we have mainly concentrated on tunneling method by using Proca equation of massive spin-$1$ field to find the tunneling probability of outgoing particles and the Hawking temperature at the outer event horizon of the neutral spinning dilaton black hole. Again we have derived Hawking temperature for general dilaton coupling. By tunning that coupling at certain values we have been able to show the radiation temperature of a particular class of black holes like $BTZ$ which is consistent with the result derived in \cite{CZH} in the context of Hawking radiation. It would also be interesting to study further the tunneling of other type of spin particles across the horizon of the black hole considered here.

\noindent{\bf{Acknowledgments:}}  
I would like to thank R. Uniyal for providing some references.

\end{document}